\documentclass[a4paper, 12pt,psfig]{iopart}
\usepackage{iopams}
\usepackage{epsf}
\usepackage{euscript}
\usepackage{amsthm}
\usepackage{setstack}
\usepackage{graphicx}
\usepackage{color}
\usepackage{bm}

\makeatletter

\newcommand{\Rmnum}[1]{\expandafter\@slowromancap\romannumeral #1@}
\makeatother

\begin{document}
\title{Isospectral graphs with identical nodal counts}
\author{Idan Oren$^{1}$ and Ram Band$^{1,2}$}
%\today
\address{$^{1}$Department of Physics of Complex Systems,
Weizmann Institute of Science, Rehovot 76100, Israel.}
\address{$^{2}$School of Mathematics, University of Bristol, Bristol BS8 1TW, UK}
 \ead{\mailto{idan.oren@weizmann.ac.il}}  \ead{\mailto{rami.band@bristol.ac.uk}}

\begin{abstract}
According to a recent conjecture, isospectral objects have
different nodal count sequences \cite{GnutSmilSond}. We study
generalized Laplacians on discrete graphs, and use them to construct
the first non-trivial counter-examples to this conjecture.\\
In addition, these examples demonstrate a surprising connection
between isospectral discrete and quantum graphs.
\end{abstract}

\section{Introduction}
  \label{sec:intro}
Nodal structures on continuous manifolds have been investigated ever
since the days of Chladni. His work was experimental and involved
the observation of nodal lines on vibrating plates. His research was
resumed on a more rigorous footing by the pioneering works of Sturm
\cite{Sturm1,Sturm2,Hinton},
Courant \cite{Cou53} and Pleijel \cite{Pleijel}.\\
In recent years a surge of research has begun on inverse nodal
problems, i.e. learning about the geometry of a system by observing
its nodal features \cite{GNKASM06,nodaltrace-wittenberg, KASM08, KL09, BandOrenSmil}.
This research follows what is already known for
many years in the regime of inverse spectral problems: one can
deduce geometrical information about a system by observing its
spectrum.\\
A key question in the framework of inverse spectral theory was posed
by Mark Kac who asked (1966):``can one hear the shape of a drum?"
\cite{Kac}. Generally speaking, this question raises the issue of
whether this information is unique. In other words, are there
non-congruent systems with the exact same spectrum? (these are
called isospectral systems). It turns out that the answer to this
question is positive. Milnor was the first to show that there are
isospectral systems in the case of flat tori in 16 dimensions
\cite{Milnor}. After him we should mark a few names who contributed
significantly to the study of the subject: Sunada \cite{Sunada}
(Riemannian manifolds), Gordon ,Webb and Wolpert \cite{GorWebbWol}
as well as Buser \textit{et al.} \cite{BCDS94} (domains in
$\mathbb{R}^2$), Band \textit{et al.} \cite{BandParBen} (quantum
graphs) and Godsil and McKay \cite{GodMcK} as well as
Brooks \cite{B99} (discrete graphs).\\
As a matter of fact, in the context of graphs, G\"{u}nthard and
Primas \cite{GunPrim} preceded Kac, raising the same question
regarding the spectra of graphs with relation to H\"{u}ckel's theory
(1956). A year later Collatz and Sinogowitz presented the first pair
of isospectral trees \cite{CollSinog}.

As mentioned, aside from the spectrum, one can also try to mine
information from the eigenfunctions of a given system. Today, it is
known that there exists geometrical information in the nodal
structures and nodal domains of eigenfunctions of manifolds,
billiards and graphs
\cite{GNKASM06,nodaltrace-wittenberg,KASM08,KL09}. Furthermore, it
is known that this geometrical information is different from the
information one can deduce solely by observing the spectrum. The
pioneering work began with Gnutzmann \textit{et al.}
\cite{GnutSmilSond,GnutKarSmil}, and continued with many other
papers, such as \cite{Klawonn} for example.

In particular, Gnutzmann \textit{et al.} \cite{GnutSmilSond}
conjectured that isospectral systems could be differentiated
by their nodal domain counts (we shall refer to it simply
as the `conjecture' throughout the paper). This conjecture has
proven to be quite a strong one with many numerical and analytical
evidence to back it up. In particular, in the case of graphs, both
quantum and discrete there exist much numerical evidence as well as
rigorous proofs for the validity of the aforementioned conjecture,
see for example \cite{BandShapSmil,Oren, BandOrenSmil}. In addition, the conjecture was proven to hold for a family of isospectral four dimensional tori \cite{Klawonn}. However, it was found recently that for a different method of counting, there exist a family of isospectral pairs of flat tori, sharing the same nodal domain counts  \cite{BrueningFajman}. This serves as a first counter-example to the conjecture.

In this paper, we would like to focus on the conjecture within the
context of discrete graphs. We shall first demonstrate its strength
and present some known results. Our main topic, however, is to
display the first counter-example to the conjecture. To this end we
will need to broaden our view from the usual operators defined on
graphs, to the more general setting of weighted graphs.

In addition we would like to report a peculiarity which involves the
discrete graphs of the counter-example. It turns out that this pair
of isospectral (discrete) graphs are also isospectral as quantum
graphs. This is intriguing since we have not been able to underatsnd
this phenomena, nor could we build this isospectral pair using any
of the (many) known methods which produce isospectral quantum
graphs.

\subsection{Discrete nodal domain theorems}
  \label{subsec:nd theorems}
Sturm \cite{Sturm1,Sturm2,Hinton}, and Courant \cite{Cou53} after
him, were the first to give analytical results about nodal domain
counts on continuous systems. Denoting the nodal count sequence by
$\{ \nu_n \}$, Courant's nodal domain theorem can be generally
phrased as $\nu_n\le n$.\\
In 1950 Gantmacher and Krein \cite{GantKrein} investigated the sign
patterns of eigenvectors of tridiagonal graphs, and in the 1970's
Fiedler wrote a couple of papers about the sign pattern of
eigenvectors of acyclic matrices (matrices which are defined on
trees) \cite{Fiedler1,Fiedler2}. Both Gantmacher and Krein, as well
as Fiedler did not formulate their findings in the language of nodal
domains. It took almost thirty years for the discrete counterpart of
the Courant nodal domain theorem to appear. Gladwell \textit{et al.}
\cite{GladZhu} and Davies \textit{et al.} \cite{Davies} were the
first to discover this analogue, and soon afterwards they were
followed by Biyikoglu \cite{Biyi} (who formulated a nodal domain
theorem for trees). Recently a lower bound for the nodal count was
derived by Berkolaiko \cite{Berko}. This bound is given explicitly
by $n-l \le \nu_n$, where $l$ is the number of independent cycles of
the graph.

Trees are an extremal class of graphs in the sense that for a given
number of vertices, they are the smallest connected graph (least
number of edges). For trees, assuming some generic conditions (which
are manifested by the fact that the eigenvectors do not vanish on
any of the vertices), it was proven that the nodal domain count of
the $n^{th}$ eigenvector of the Laplacian matrix has exactly $n$
nodal domains \cite{Biyi,Berko}. Therefore all trees (with the same
number of vertices) share the same nodal domain count sequence.
Furthermore, it is known that almost all tree graphs are isospectral
\cite{Schwenk} (meaning that almost any tree has a isospectral
mate). This means that we cannot resolve the isospectrality using
nodal domain counts, when it comes to trees. This shortcoming of the
conjecture is well known, and to the best of our knowledge, occurs
only for trees.\\
If we introduce weighted graphs, then there exist two more trivial
counter-examples: complete graphs and polygon graphs (connected
graphs in which all vertices have degree 2). In the case of complete
weighted graphs, the first eigenvector has only one nodal domain and
all other eigenvectors have exactly 2 nodal domains. Hence, they are
an obvious counter-example. It should be noted that complete graphs
are also extremal in the sense that for a given number of vertices,
they are the largest connected graph (largest number of edges). For
polygons, it can be shown (using the Courant bound \cite{Cou53} and
Berkolaiko's bound \cite{Berko}) that polygons always have the same
nodal count.

As far as the authors know, these three cases are the only
counter-examples to the conjecture.

Aside from this extreme cases, in all isospectral graph pairs which
were compared (analytically and numerically), different nodal domain
sequences were observed \cite{OrenSmil}. In addition we have a proof
for the conjecture for a certain class of discrete graphs
\cite{Oren}.

Up until now, we only discussed isospectrality of the traditional
matrices defined on graphs, most notably the adjacency matrix and
the Laplacian. Additional work was done on less studied matrices
such as
the signless Laplacian and the normalized Laplacian.\\
However, since nodal domain theorems were proven for a more general
class of matrices (generalized Laplacians), it is natural to test
the conjecture for this class as well.

The paper is organized as follows. We will begin with some
background and necessary definitions. The following section will
describe the method of construction of isospectral weighted graphs.
Then we will present the counter-example to the conjecture and
finally prove the isospectrality of the quantum analogue of our
discrete graphs.

\section{Definitions}
  \label{sec:def}

\subsection{Discrete graphs}
  \label{subsec:discrete graphs}
A \textit{graph} $G$ is a set $\mathcal{V}$ of vertices connected by
a set $\mathcal{E}$ of edges. The number of vertices is denoted by
$V = |\mathcal{V}|$ and the number of edges is $E=|\mathcal{E}|$.
The \textit{degree} (\textit{valency}) of a vertex is the number of
edges which are connected to it. The number of independent cycles of
a graph is denoted by $l$ and is given by $l=E-V+C$, where $C$ is
the number of connected components of the graph.

The \textit{weighted adjacency matrix} (\textit{connectivity}) of
$G$ is the symmetric $V\times V$ matrix $A=A(G)$ whose entries are
given by:
\begin{displaymath}
A_{ij}=\left\{ \begin{array}{ll} w_{ij}, & \textrm{if $i$ and $j$
are adjacent}\\
0, & \textrm{otherwise}
\end{array} \right.
\end{displaymath}

The $w_{ij}$'s values are called weights and are usually taken to be
positive. For non-weighted graphs, all the weights are equal to
unity. A diagonal element in $A$ corresponds to a loop, which is an
edge connecting a vertex to itself. We shall only discuss graphs
without loops.\\

A \textit{generalized Laplacian}, $L(G)$, also known as a
\textit{Schr\"{o}dinger operator} of $G$, is a matrix
\begin{displaymath}
L_{ij}=\left\{ \begin{array}{ll} -w_{ij}, & \textrm{if $i$ and $j$
are adjacent}\\
P_i, & \textrm{if $i=j$}\\
0, & \textrm{otherwise}
\end{array} \right.
\end{displaymath}
where $P_i$ is an arbitrary on-site potential which can assume any
real value and $w_{ij}>0$. The \textit{combinatorial Laplacian}
results by taking all weights to be unity, and $P_i =
-\sum_{j}w_{ij} = -deg(i)$, where $deg(i)$ is the degree of the
vertex $i$. This way, the sum of
each row, or column is equal to zero.\\
The eigenvalues of $L(G)$ together with their multiplicities, are
known as the \textit{spectrum} of $G$. To the $n^{th}$ eigenvalue,
$\lambda_n$, corresponds (at least one) eigenvector whose entries
are labeled by the vertex indices, i.e.,
$\phi_n=(\phi_n(1),\phi_n(2),\ldots,\phi_n(V))$. A \textit{nodal
domain} is a maximally connected subgraph of $G$ such that all
vertices have the same sign with respect to $\phi_n$. The number of
nodal domains of an eigenvector $\phi_n$ is called a \textit{nodal
domain count}, and will be denoted by $\nu_n$. The \textit{nodal
count sequence} of a graph is the number of nodal domains of
eigenvectors of the Laplacian, arranged by increasing eigenvalues.
This sequence will be denoted by $\{ \nu_n
\}_{n=1}^{V}$.\\
We recall that the known bounds for the nodal count
\cite{GladZhu,Davies,Berko} are
\begin{equation}
n-l \le \nu_n \le n.
\end{equation}

\subsection{Quantum graphs}
  \label{subsec:quantum graphs}
To define \textit{quantum graphs} a metric is associated to     %%%%%% RAMI: Defining Quantum Graphs %%%%%%%%%%%%%
$G$. That is, each edge is assigned a positive length: $L_e \in
(0,\infty)$. The total length of the graph will be denoted by
$\mathcal{L}=\sum_{e\in \mathcal{E}} L_E$. This enables to define
the metric Laplacian (or Schr\"odinger) operator on the graph as the
Laplacian in 1-d $-\frac{{\rm d}^2 \ }{{\rm d} x^2}$ on each bond.
The domain of the Schr\"odinger operator on the graph is the space
of functions which belong to the Sobolev space $H^2(e)$ on each edge
$e$ and satisfy certain vertex conditions. These vertex conditions
involve vertex values of functions and their derivatives, and they
are imposed to render the operator self adjoint. We shall consider
in this paper only the so-called Neumann vertex conditions:
\begin{equation}
{\rm Neumann}\ \ \ \ \forall v \  \  : \ \ \  \sum_{e\in S^{(v)}}
\left. \frac{{\rm d}\ \ \ }{{\rm d} x_{e} }\ \psi_{e}(x_{e})\right
|_{x_{e}=0} \ = \ 0\ \
 \label{eq:boundary_cond}
\end{equation}
where $S^{(v)}$ is the set of all edges connected to the vertex $v$.
The derivatives in (\ref{eq:boundary_cond}) are directed out of the
vertex $v$.  The eigenfunctions are the solutions of the edge
Schr\"odinger equations:
\begin{eqnarray}
\ \ & & \ \ \forall e\in E \  \ -\frac{{\rm d}^2 \ }{{\rm d} x^2}
\psi_e=k^2\psi_e ,
 \label{eq:schrodinger_eq}
\end{eqnarray}
which satisfy at each vertex the Neumann conditions
(\ref{eq:boundary_cond}). The spectrum $\{k_n^2\}_{n=1}^{\infty}$ is
discrete, non-negative and unbounded. One can generalize the
Schr\"odinger operator by including potential and magnetic flux
defined on the bonds.  Other forms of vertex conditions can also be
used. However, these generalizations will not be addressed here, and
the interested reader is referred to two recent reviews
\cite{GS06,Kuch08}.

Finally, Two graphs, $G_1$ and $G_2$, are said to be
\textit{isospectral} if they posses the same spectrum (same
eigenvalues with the same multiplicities). In perfect analogy, two
graphs with the same nodal domain sequence will be referred to as
\textit{isonodal}. These two definitions hold both for discrete and
quantum graphs.

\section{Isospectrality and isonodality}
  \label{sec:Iso+Iso}

\subsection{Isospectral graphs construction}
  \label{subsec:Iso}

Our method for constructing isospectral graphs is a variation of a
method described in \cite{MM03}, called the \textit{line graph
construction}. This method uses the gallery of isospectral billiards
of Buser \textit{et al.} \cite{BCDS94} in order to build isopectral
discrete graphs.
A similar idea was used by Gutkin \textit{et al.} \cite{GutSmil}   %%%%% CHECK IF THAT'S TRUE About Brooks %%%%%%
to construct isospectral discrete and quantum graphs.\\
A line graph is built from a ``parent" graph in the following way:
each edge becomes a vertex, and two vertices in the line graph are
adjacent if and only if their corresponding edges shared a vertex in
the parent graph. In \cite{MM03} an example is given, based on the
first family of isospectral domains in \cite{BCDS94} called the
$7_{1}$ family. Our method is simpler than the one in \cite{MM03}.
It results with graphs with the same topology as in \cite{MM03}, but
with different Laplacian matrices.

\begin{figure}[h]
 \centering
 \scalebox{0.6}{\includegraphics{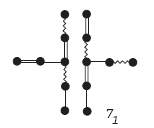}}
 \caption{The $7_1$ family in the representation presented in \cite{OS01}.}
   \label{fig:7-1 pair as graphs}
\end{figure}

Instead of using the gallery of billiards as it appears in
\cite{BCDS94}, we use a graph representation of them as it is
described in \cite{OS01}. In particular, the $7_1$ family is shown
in figure \ref{fig:7-1 pair as graphs}.

\begin{figure}[h]
  \centering
 \scalebox{0.5}{\includegraphics{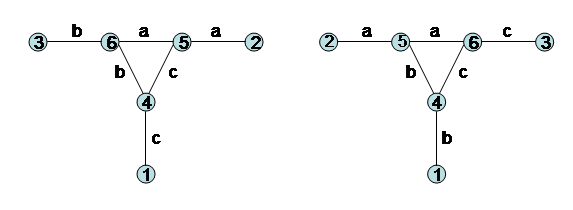}}
 \caption{Two isospectral graphs constructed through the line graph construction from the $7_1$ billiards.}
  \label{fig:7-1 line graphs}
\end{figure}

We consider the two graphs in figure \ref{fig:7-1 pair as graphs} as
parent graphs and apply the line graph construction on them. We
still have to specify how we assign weights in the resulting line
graphs. We start by assigning three different weights: $a,b,c>0$ to
each of the three types of edges in the parent graphs. Suppose that
in the parent graph, an edge of weight $a$ shared a vertex with an
edge of weight $b$ ($a\neq b$). Then, in the line graph, the
corresponding vertices
would be connected by an edge of the remaining weight $c\neq a,b$.\\
The two resulting weighted line graphs are shown in figure
\ref{fig:7-1 line graphs}. Let us denote the left graph by $G_1$ and
the right one by $G_2$.
The generalized Laplacians of the two graphs are given explicitly by
following matrices:

\begin{eqnarray}
L_1 = - \left( \begin{array}{cccccc} %%% How to put these 2 matrices side by side? NUMBERING%%%
0 & 0 & 0 & c & 0 & 0 \\
0 & 0 & 0 & 0 & a & 0 \\
0 & 0 & 0 & 0 & 0 & b \\
c & 0 & 0 & 0 & c & b \\
0 & a & 0 & c & 0 & a \\
0 & 0 & b & b & a & 0 \\
\end{array} \right)
  \label{Dirichlet Laplacian 1}
L_2 = - \left(
\begin{array}{cccccc}
0 & 0 & 0 & b & 0 & 0 \\
0 & 0 & 0 & 0 & a & 0 \\
0 & 0 & 0 & 0 & 0 & c \\
b & 0 & 0 & 0 & b & c \\
0 & a & 0 & b & 0 & a \\
0 & 0 & c & c & a & 0 \\
\end{array} \right)
  \label{Dirichlet Laplacian 2}
\end{eqnarray}

It is not hard to check that for any $a,b,c$, the characteristic
polynomials of $L_1$ and $L_2$ are identical and hence the graphs
are isospectral. Another way to prove the isospectrality is to
construct the transplantation matrix $T$ such that $T^{-1}L_1T =
L_2$. Then it is clear that the two matrices are similar and
therefore isospectral. The transplantation matrix between $L_1$ and
$L_2$ is
\begin{equation}
T =  \left( \begin{array}{cccccc}
0 & -1 & 0 & 0 & 0 & 1 \\
-1 & 0 & 0 & 0 & 1 & 0 \\
0 & 0 & -1 & 1 & 0 & 0 \\
0 & 0 & 1 & 1 & 0 & 0 \\
0 & 1 & 0 & 0 & 0 & 1 \\
1 & 0 & 0 & 0 & 1 & 0 \\
\end{array} \right)
  \label{transplantation matrix 7_1 discrete}
\end{equation}

\noindent The same construction can be carried out for any graph in
the gallery of \cite{OS01}.\\
We can construct many more isospectral graphs by using polynomials
in $L_1$ and $L_2$. Namely, for any polynomial $P$, we will consider
$P(L_1)$ and $P(L_2)$ as the Laplacian matrices of two new weighted
graphs (assuming that $P(L_1)$ and $P(L_2)$ are indeed generalized
Laplacians as defined in section \ref{subsec:discrete graphs}). These
two graphs might be topologically different than the original $G_1$
and $G_2$. Since we have a transplantation matrix, it is clear that
$P(L_1)$ and $P(L_2)$ are similar matrices and therefore the
resulting graphs are also isospectral.

\subsection{Failure of the conjecture regarding nodal domain counts}

We have introduced the conjecture that the isospectrality between
graphs can be resolved by counting nodal domains. We have also said
that three known cases (trees, polygons and complete graphs) are
exceptions to this conjecture. We now prove that  %%%%%CHECK this statement and prove it by Courant+Berkolaiko %%%%%%
$G_1$ and $G_2$ cannot be resolved by counting nodal domains. This
is a non-trivial exception to the conjecture.

We define the vertices with degree larger than one as the
\textit{interior vertices} (vertices $4,5,6$), and the rest as
\textit{boundary vertices} (vertices $1,2,3$).\\
We begin by checking the relations between the interior and boundary
vertices.\\
Let $\phi_n^1$ be the $n^{th}$ eigenfunction of $L_1$, and
$\phi_n^2$ be the $n^{th}$ eigenfunction of $L_2$. For the first
graph, we get the following relations:

\begin{equation}
\phi_n^1(1) = \frac{-c}{\lambda_i} \phi_n^1(4)\\
\phi_n^1(2) = \frac{-a}{\lambda_i} \phi_n^1(5)\\
\phi_n^1(3) = \frac{-b}{\lambda_i} \phi_n^1(6)
  \label{eq:relation between interior and boundary vertices}
\end{equation}
For the second graph, we get the same relations with $b$ and $c$
replaced. Therefore, since the weights are positive, if
$\lambda_n<0$ then each boundary vertex has the same sign as the
interior vertex connected to it. This means that for $\lambda_n<0$,
the boundary vertices will not contribute to the nodal domain count.
On the other hand, if $\lambda_n>0$, each boundary vertex has an
opposite sign than the interior vertex connected to it. This means
that for $\lambda_n>0$, the boundary vertices will contribute three
to the nodal domain count. The most important point is that the
contribution of the boundary vertices to the nodal count depends on
the spectrum, and since the two graphs are isospectral, it is the
same for both graphs. As a result, it is enough to compare only the
nodal
count sequence of the interior vertices.\\
The interior vertices form a triangle. Therefore the nodal domain
count of any vector, on the subgraph induced by the interior
vertices, is either one or two. By computing the rest of the
equations, and with the aid of (\ref{eq:relation between interior
and boundary vertices}), we can formulate the conditions for having one or two nodal domains.\\
The interior nodal domain count of a vector $\phi_i$ (of any of the
graphs) is one if and only if one of the following is true:

\begin{eqnarray}
\{\lambda_n<0\  \textrm{and}\  \left| \lambda_n \right|>\max{(a,b,c)}\}\ \textrm{or} \nonumber \\
\{\lambda_n>0\  \textrm{and}\  \lambda_n<\min{(a,b,c)}\}.
  \label{eq:cases when nd count is one}
\end{eqnarray}
In any other case, the nodal domain count is two.\\
In other words, a nodal domain count of a specific vector, is
determined uniquely by the corresponding eigenvalue. Therefore, the
entire nodal domain counts of the graphs are determined by the
spectrum, and since the two graphs are isospectral, the nodal
count sequence does not resolve the isospectrality.\\

\begin{figure}[h]
  \centering
 \scalebox{0.5}{\includegraphics{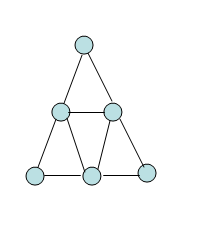}}
 \caption{The graph obtained by applying a polynomial $P(x)=cx^2$.}
  \label{fig:second order graph}
\end{figure}

\begin{figure}[h]
  \centering
 \scalebox{0.5}{\includegraphics{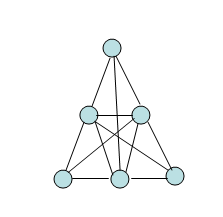}}
 \caption{The graph obtained by applying a polynomial $P(x)=c_1x+c_2x^2$}
  \label{fig:First and second order graph}
\end{figure}

As we have shown in subsection \ref{subsec:Iso}, for any polynomial
$P$, the two graphs represented by $P(L_1)$ and $P(L_2)$ are
isospectral. We will now show that these graphs are also isonodal,
thus extending our family of counter-examples to the conjecture.\\
Assuming that the weights $a,b,c$ are rationally incommensurate, the
following observations can be easily proven:

\begin{itemize}
\item{If the polynomial consists only of a second degree term
($P(x)=cx^2$), then the obtained graphs $P(L_1)$ and $P(L_2)$ are
given by figure \ref{fig:second order graph}.}

\item{If $P(x)=c_1x+c_2x^2$, then the obtained graphs $P(L_1)$ and $P(L_2)$
are given by figure \ref{fig:First and second order graph}.}

\item{Polynomials of third degree or larger represent weighted,
complete graphs, which are trivial counter-examples to the
conjecture (see section \ref{subsec:nd theorems}).}
\end{itemize}

For these reasons, we only need to check the first two cases
above.\\
The resulting graphs, In both cases, clearly have the same
eigenvectors as $G_1$ and $G_2$. Then, using
\eref{eq:relation between interior and boundary vertices} and
\eref{eq:cases when nd count is one}, it can be easily shown that in
both types of graphs, the nodal count is determined by the spectrum,
precisely as
is the case with $G_1$ and $G_2$.\\
We conclude that both types of graphs are isonodal and as a
consequence, they are also non-trivial counter-examples to the
conjecture.

\emph{Remark:} Applying the line graph construction to other
families from the gallery in Buser \textit{et al.} \cite{BCDS94},
one can build many pairs of isospectral graphs. Some of these pairs
are isonodal (such as the $7_2$ and $7_3$ families) and some are not
(such as the $13_2$ family).

\section{Isospectral quantum graphs}

When we come to discuss quantum graphs, we need to define the
lengths of the different edges. The weights we put on the weighted
discrete graphs can be viewed as coupling constants. Thus, the most
intuitive notion is to associate a length which is inversely
proportional to the weights. If we also specify the vertex
conditions, we go from the realm of discrete graphs into the realm
of quantum graphs.\\
We then come to ask the following interesting question: is the
isospectrality preserved when we enter the world of quantum graphs?
This question is only a small part of a much broader subject - the
spectral relations between quantum graphs and the underlying
discrete graphs. This subject was addressed by several authors in
the past, see for example \cite{vB85,vB01,C97,P}. However, most of
these references have a complete analysis only for equilateral
quantum graphs, with Neumann vertex conditions. The graphs $G_1$ and
$G_2$ are \emph{not} equilateral, and therefore we cannot make an
a-priori prediction whether or not the isospectrality is preserved.

\noindent Nevertheless, we will show by direct computation that
$G_1$ and $G_2$ are indeed isospectral as quantum graphs, once
Neumann vertex conditions are considered at all vertices.\\
\noindent A function $\psi$ on the graph, which is continuous on the
vertices, can be written as:
\begin{equation}
\psi|_{(i,j)}=\frac{1}{\sin{kL_{ij}}}\left[\phi(i)
\sin{k(L_{ij}-x)}+\phi(j)\sin{kx}\right],
  \label{Wave function}
\end{equation}
where $\phi(i)$ is the value of the function on the vertex $i$ and
$L_{ij}$ is the length of the edge $(i,j)$. Note, that we still use
the notations $a,b,c$ to denote the lengths of the edges.\\

The Neumann vertex conditions on the boundary vertices for $G_1$,
dictate these relations:

\begin{equation}
\phi(1)=\frac{\phi(4)}{\cos{kc}}\\
\phi(2)=\frac{\phi(5)}{\cos{ka}}\\
\phi(3)=\frac{\phi(6)}{\cos{kb}}
  \label{Neumann graph 1}
\end{equation}
The Neumann conditions on the interior vertices are (we make use of
(\ref{Neumann graph 1})):

\begin{equation}
\hspace{-3cm}
-\frac{1}{\sin{kc}}\left[-\frac{\phi(4)}{\cos{kc}}+\phi(4)\cos{kc}\right]+\frac{1}{\sin{kc}}\left[-\phi(4)\cos{kc}+\phi(5)\right]
+\frac{1}{\sin{kb}}\left[-\phi(4)\cos{kb}+\phi(6)\right]=0
  \label{Kirchoff graph 1:1}
\end{equation}
\begin{equation}
\hspace{-2.7cm}
\frac{1}{\sin{ka}}\left[-\frac{\phi(5)}{\cos{ka}}+\phi(5)\cos{ka}\right]+\frac{1}{\sin{kc}}\left[\phi(5)\cos{kc}-\phi(4)\right]
-\frac{1}{\sin{ka}}\left[-\phi(5)\cos{ka}+\phi(6)\right]=0
  \label{Kirchoff graph 1:2}
\end{equation}
\begin{equation}
\hspace{-2.64cm}
\frac{1}{\sin{kb}}\left[-\frac{\phi(6)}{\cos{kb}}+\phi(6)\cos{kb}\right]+\frac{1}{\sin{kb}}\left[\phi(6)\cos{kb}-\phi(4)\right]
+\frac{1}{\sin{ka}}\left[\phi(6)\cos{ka}-\phi(5)\right]=0
  \label{Kirchoff graph 1:3}
\end{equation}
This can be written more conveniently as a matrix-vector product:
\begin{equation}
A^1(k)\bm{\phi}=0
  \label{Matrix of secular equation}
\end{equation}
Where $1$ comes to represent that this is the matrix corresponding
to $G_1$, and $\bm{\phi} = (\phi(4),\phi(5),\phi(6))$.\\
The matrix $A^1(k)$ is:

\[ A^1(k) =  \left( \begin{array}{ccc}
2\cot{2kc}+\cot{kb} & \frac{-1}{sin{kc}} & \frac{-1}{sin{kb}} \\
\frac{-1}{sin{kc}} & 2\cot{2ka}+\cot{kc} & \frac{-1}{sin{ka}} \\
\frac{-1}{sin{kb}} & \frac{-1}{sin{ka}} & 2\cot{2kb}+\cot{ka} \\
\end{array} \right)\]
(\ref{Matrix of secular equation}) has a solution if and only if

\begin{equation}
h^1(k)\equiv\det{A^1(k)}=0
  \label{secular equation}
\end{equation}
$h^1(k)$ is called the \textit{secular function}, and equation
(\ref{secular equation}) is called the \textit{secular equation}. It
is fulfilled at the values $k$ which are in the spectrum of the
Laplacian of the graph. We can get $A^{2}(k)$ by switching the
lengths $b$ and $c$ in $A^1(k)$. It can be easily checked that
$h^1(k) = h^{2}(k)\equiv h(k)$, hence the graphs are isospectral.\\

Although we have proven that $G_1$ and $G_2$ are isospectral as
quantum graphs, the profound reason for this is still a riddle for
us. The recent papers on isospectrality \cite{BandParBen,ParBand}
generalize former seminal papers such as those of Sunada
\cite{Sunada} and Buser \textit{et al.} \cite{BCDS94}, and can
produce many of the known examples of isopectral quantum graphs.
However, we were not able to build the two graphs $G_1$ and $G_2$
using the constructions described in \cite{BandParBen,ParBand}.
Furthermore, We were unable to build a transplantation matrix for
the quantum graphs (although there is a transplantation matrix for
the discrete case - see (\ref{transplantation matrix 7_1
discrete})). It should be emphasized that all isospectral quantum
graphs which are built using any of the methods in
\cite{Sunada,BCDS94,BandParBen,ParBand} posses a transplantation
matrix between the two graphs. In \cite{BandSawSmil}, the authors
consider the two graphs in the present paper and turn them into
scattering systems. They prove that there is no transplantation
which involves the values of the eigenfunctions on the vertices.
They do not, however, eliminate the possibility of having any other
form of transplantation. All these pieces of evidence suggest that
$G_1$ and $G_2$ might belong to a new class of isospectral quantum
graphs.

\emph{Remark:} Unlike the graphs $G_1$ and $G_2$ which correspond
the the $7_1$ family, the isospectrality is not preserved in the
$7_2$ and $7_3$ graphs (i.e., the corresponding quantum graphs are
not isospectral). This leads us to contemplate the issue of
converting isospectral weighted discrete graphs into their
isospectral quantum analogues. How to do so, or whether at all it is
possible, remains an open problem.

\section{Summary}
  \label{sec:summary}

The conjecture that isospectrality can be lifted by comparing nodal
domain counts was originally stated for flat tori of dimension
larger than three \cite{GnutSmilSond}. Later on, this conjecture was
proven for four-dimensional flat tori \cite{Klawonn}. However, using a different counting method, a family of both isosepctral and isonodal pairs of flat tori was discovered \cite{BrueningFajman}.

The conjecture was imported into the realm of graphs where it
was proven for some quantum and discrete graphs \cite{BandShapSmil,
Oren}. In addition, there exist much numerical evidence for the
validity of the conjecture in discrete graphs \cite{OrenSmil} (using
a construction by Godsil and McKay \cite{GodMcK}).\\

In this paper we show that for discrete graphs, the conjecture is
not true in its most general form. What we demonstrate is that if we
use generalized Laplacians, the conjecture ceases to be valid even
for graphs which are not extremal. One should keep in mind that if
we restrict ourselves only to the traditional matrices - the
adjacency and Laplacian matrices - then the only known
counter-examples to the conjecture are trees.

The paper also presents an intriguing connection between isospectral
discrete and quantum graphs. The fact that both the discrete graphs,
and their quantum analogues are isospectral, calls for more study on
the relation between these two regimes.

\section{Acknowledgements}
  \label{sec:acknowledgements}
The authors warmly thank Uzy Smilansky for his continuous support, significant encouragement and for many invaluable discussions. The work was supported by ISF grant 169/09. RB is supported by EPSRC, grant number EP/H028803/1.

%{10}


\section*{References}
\begin{thebibliography}{10}

\bibitem{Sturm1} C. Sturm, \textit{M'moire sur les \'{e}quations diff\'{e}rentielles lin\'{e}aires du second ordre},
J. Math. Pures Appl., 1:106�186, 1836.

\bibitem{Sturm2}  C. Sturm, \textit{M`moire sur une classe d�\'{e}quations \`{a} diff\'{e}rences partielles},
J. Math. Pures Appl., 1:373�444, 1836.

\bibitem{Hinton} D. Hinton, \textit{Sturm�s 1836 oscillation results: evolution of the
theory}, In Sturm-Liouville theory, pages 1�27. Birkh�user, Basel,
2005.

\bibitem{Cou53} R. Courant and D. Hilbert, \textit{Methods of
Mathematical Physics}, Vol. 1, Interscience, New York, 1953.

\bibitem{Pleijel} A. Pleijel, \textit{Remarks on Courant's nodal line theorem}, Comm.
Pure Appl. Math. 9, 543-550, 1956.

\bibitem{Kac} M. Kac, \textit{Can one hear the shape of a drum?}, American
Mathematical Monthly  73 (4, part 2), 1�23, 1966.

\bibitem{Milnor} J. Milnor, \textit{Eigenvalues of the Laplace operator on
certain manifolds}, Proceedings of the National Academy of Sciences
of the United States of America 51, 542ff, 1964.

\bibitem{Sunada} T. Sunada, \textit{Riemannian coverings and isospectral
manifolds}, Ann. Of Math. (2) 121 (1), 169�186, 1985.

\bibitem{GorWebbWol} C. Gordon, D. Webb, S. Wolpert, \textit{One
Cannot Hear the Shape of a Drum}, Bulletin of the American
Mathematical Society 27 (1), 134�138, 1992.

\bibitem{BCDS94} P. Buser, J. Conway, P. Doyle and K. D. Semmler,
\textit{Some planar isospectral domains}, International Mathematics
Research Notices 9: 391ff' 1994.

\bibitem{BandParBen} R. Band, O. Parzanchevski and G. Ben-Shach, \textit{The isospectral
fruits of representation theory: quantum graphs and drums}, J. Phys.
A: Math. Theor. 42 175202, 2009.

\bibitem{GodMcK} C. D. Godsil and B. D. McKay, \textit{Constructing cospectral
graphs},  Aequationes Mathematicae 25, 257-268, 1982.

\bibitem{B99} R. Brooks, \textit{Non-Sunada graphs}, Ann. Inst. Fourier 49
707�25, 1999.

\bibitem{GunPrim} Hs. H. Gunthard and H. Primas, \textit{Zusammenhang von Graphentheorie und
MO-Theorie von Molekeln mit Systemen konjugierter Bindungen}, Helv.
Chim. Acta 39, pp. 1645�1653, 1956.

\bibitem{CollSinog} L. Collatz and U. Sinogowitz, \textit{Spektren endlicher Grafen}, Abh.
Math. Sem. Univ. Hamburg.  21, pp. 63�77, 1957.

\bibitem{GNKASM06} S. Gnutzmann, P.D. Karageorge, U. Smilansky, \textit{Can
One Count the Shape of a drum?}, Phys. Rev. Lett. \textbf{97},
090201 (2006).

\bibitem{nodaltrace-wittenberg} S. Gnutzmann, P.D. Karageorge, U. Smilansky,
\textit{A trace formula for the nodal count sequence -- Towards
counting the shape of separable drums}, Eur. Phys. J. Special Topics
\textbf{145}, 217 (2007)

\bibitem{KASM08} P.D. Karageorge, U. Smilansky, \textit{Counting
nodal domains on surfaces of revolution}, J. Phys. A \textbf{41},
205102 (2008).

\bibitem{KL09} D. Klawonn, \textit{Inverse Nodal Problems}, J. Phys.
A \textbf{42}, 175209 (2009).

\bibitem{GnutSmilSond} S. Gnutzmann, U. Smilansky and N. Sondergaard, \textit{Resolving
isospectral `drums' by counting nodal domains}, J. Phys. A: Math.
Gen. 38 8921�33, 2005.

\bibitem{GnutKarSmil} S. Gnutzmann, P. D. Karageorge and U. Smilansky, \textit{Can One Count the
Shape of a Drum?}, Phys. Rev. Lett. 97 090201, 2006.

\bibitem{Klawonn} J. Br\"{u}ning, D. Klawonn, and C. Puhle,
\textit{Remarks on "Resolving isospectral `drums' by counting nodal
domains"}, J. Phys. A, 40(50):15143-15147, 2007.

\bibitem{BandShapSmil} R. Band, T. Shapira and U. Smilansky, \textit{Nodal domains on isospectral
quantum graphs: the resolution of isospectrality?}, J. Phys. A.:
Math. Gen. 39 13999-4014, 2006.

\bibitem{Oren} I. Oren, \textit{Nodal domain counts and the chromatic
number of graphs}, J. Phys. A: Math. Theor. 40 9825, 2007.

\bibitem{BandOrenSmil} R. Band, I. Oren, and U. Smilansky, \textit{Nodal domains on graphs - How
to count them and why?}, in \cite{EKKST}.

\bibitem{BrueningFajman} J. Br\"{u}ning, D. Fajman, \textit{On the nodal count for flat tori}, to appear in Comm. Math. Phys., 2011.

\bibitem{GantKrein} F. R. Gantmacher and M. G. Krein, \textit{Oscillation Matrices and
Kernels and Small Vibrations of Mechanical Systems}, State
Publishing House of Technical-Theoretical Literature, Moscow,
Leningrad, 1950. English Translation by US Atomic Energy Commission,
Washington D. C. 1961.

\bibitem{Fiedler1} M. Fiedler, \textit{Eigenvectors of acyclic matrices}, Czechoslovak Math.
J., 25, 607-618, 1975.

\bibitem{Fiedler2} M. Fiedler, \textit{A property of eigenvectors of non-negative symmetric
matrices and its application to graph theory}, Czechoslovak Math.
J., 25, 619-633, 1975.

\bibitem{GladZhu} G. M. L. Gladwell and H. Zhu, \textit{Courant�s nodal line theorem and
its discrete counterparts}' Quart. J. Mech. Appl. Math., 55(1),
1�15, 2002.

\bibitem{Davies} E. B. Davies, G. M.L. Gladwell, J. Leydold and P. F.
Stadler, \textit{Discrete Nodal Domain Theorems}, Linear Algebra and
its Applications Vol. 336, October, pp. 51-60, 2001.

\bibitem{Biyi} T. Biyiko$\breve{g}$lu, \textit{A discrete nodal domain theorem for
trees}, Linear Algebra Appl. 360, 197-205, 2003.

\bibitem{Berko} G. Berkolaiko, \textit{A lower bound for nodal count on
discrete and metric graphs}, Comm. Math. Phys., 278(3), 803�819,
2008.

\bibitem{Schwenk} A.J. Schwenk, \textit{Almost all trees are cospectral}, In: \textit{New Directions in the Theory of Graphs}, F. Harary,
Editor, Academic Press, New York, pp. 275�307, 1973.

\bibitem{GS06} S. Gnutzmann and U. Smilansky, \textit{Quantum graphs:
Applications to quantum chaos and universal spectral statistics},
Advances in Physics, 55 (5-6), 527-625, 2006.

\bibitem{Kuch08} P. Kuchment, \textit{Quantum graphs: an introduction and a brief survey},
In Analysis on graphs and its applications, volume \textbf{77} of
Proc. Sympos. Pure Math., Amer. Math. Soc., Providence, RI, 291-312
, 2008.

\bibitem{GutSmil} B. Gutkin and U. Smilansky, \textit{Can one hear the shape of a graph?},
J. Phys. A: Math. Gen. 31 6061�8, 2001.

\bibitem{MM03} P. McDonald and R. Meyers, \textit{Isospectral polygons,
planar graphs and heat content}, Proceedings of the American
mathematical society, Vol. 131, Number 11, Pages 3589-3599 S
0002-9939(03)07123-5 Article electronically published on June 18,
2003.

\bibitem{OS01} Y. Okada and A. Shudo 2001 J. Phys. A: Math. Gen. 34
5911�22.

%T. Shapira and U. Smilansky, Quantum graphs which sound the same, in
%F. Khanna and D. Matrasulov (editors), Nonlinear Dynamics and
%Fundamental Interactions, Proc. Nato Advanced Research Workshop,
%Tashkent 2004 (Springer, Berlin 2005).

%\bibitem{ShapSmil} T. Shapira and U. Smilansky, \textit{Quantum graphs which sound the
%same}, Nonlinear Dynamics and Fundamental Interactions NATO Science
%Series II: Mathematics, Physics and Chemistry vol. 213, pp 17-29,
%2005.

\bibitem{vB85} J. von Below, \textit{A Characteristic Equation
Associated to an Eigenvalue Problem on $c^2$-Networks}, Linear
Algebra and its Applications, 71: 309-325 (1985).

\bibitem{vB01} J. von Below, \textit{Can one hear the shape of a network?},
in \cite{MvBN01}, 19-36, 2001.

\bibitem{C97} C. Cattaneo, \textit{The spectrum of the continuous Laplacian on a graph},
Monatsh. Math. 124 (1997), no. 3, 215�235.

\bibitem{P} O. Post,\textit{ Equilateral quantum graphs and boundary triples}, in
\cite{EKKST}.

\bibitem{BandSawSmil} R. Band, A. Sawicki and U. Smilansky, \textit{Scattering from isospectral quantum graphs},
J. Phys. A: Math. Theor. 43 415201, 2010.

\bibitem{ParBand} O. Parzanchevski and R. Band, \textit{Linear Representations and Isospectrality with Boundary
Conditions}, J. Geom. Anal. 20 439-71, 2010.

\bibitem{Roth} J.P Roth, \textit{Le spectre du Laplacien sur un graphe Th�eorie du Potentiel} Lect Not. Math. vol.
1096 (Berlin: Springer) 521-539, 1983.

\bibitem{EKKST} P. Exner, J. P. Keating, P. Kuchment, T. Sunada, and A. Teplyaev
(Editors), \textit{Analysis on Graphs and its Applications}, Proc.
Symp. Pure Math., AMS.

\bibitem{MvBN01} F.A. Mehmeti, J. von Below, and S. Nicaise (Editors), \textit{Partial
differential equations on multistructures}, Proceedings of the
International Conference held in Luminy, April 19�24, 1999, Lecture
Notes in Pure and Applied Mathematics, 219. Marcel Dekker, Inc., New
York, 2001.

\bibitem{OrenSmil} I. Oren and U. Smilansky, private communication.



\end{thebibliography}
\end{document}